\documentclass[epjST,final]{svjour}

\usepackage{amsmath}
\usepackage{amssymb}
\usepackage{hyperref}
\usepackage{graphicx}
\usepackage{verbatim}
\usepackage{cite}

\usepackage{color}

\begin{document}

\title{Magnetic properties of commensurate Bose-Bose mixtures in one-dimensional optical lattices}
\author{  M. Dalmonte\inst{1,2}, E. Ercolessi\inst{1}, M. Mattioli\inst{1}, F. Ortolani\inst{1} and D. Vodola\inst{1}}
\authorrunning{M. Dalmonte {\it et al}.}

\institute{$^1$Dipartimento di Fisica dell'Universit\`a di Bologna and INFN, via Irnerio 46, 40126 Bologna, Italy,\\
$^2$Institute for Quantum Optics and Quantum Information of the Austrian Academy of Sciences, A-6020 Innsbruck, Austria \\
}

\date{Received: --- / Revised version: ----}

\abstract{
We investigate magnetic properties of strongly interacting bosonic mixtures confined in one
dimensional geometries, focusing on recently realized $^{87}$Rb-$^{41}$K gases with 
tunable interspecies interactions. By combining analytical perturbation theory results 
with density-ma\-trix-renormalization group calculations, we provide quantitative estimates
of the ground state phase diagram as a function of the relevant microscopic quantities, 
identifying the more favorable experimental regimes in order to access the various
magnetic phases. Finally, we qualitatively discuss the observability of such phases 
in realistic setups when finite temperature effects have to be considered.} 
\maketitle{}

\section{Introduction}
\label{sec:intro_BBmix}
Experimental advances in the preparation and manipulation of cold atoms and mole\-cules trapped in optical lattices~\cite{jaksch1998,bloch2008} have revived the theoretical interest towards models and problems that in the past played a fundamental role for the description of possible new states of matter, but lacked physical realization in solid state setups. Indeed, thanks to the possibility of using both fermionic and bosonic type of atoms and to the high tunability of the interaction shape and parameters, the phenomena that one can access with these systems are the most varied, ranging from standard superfluidity to BEC-BCS cross-over, to Mott physics and dynamical processes~\cite{bloch2008,stringari_fermi,stringari1999,lewenstein2007}.

Among the various advantages in dealing with cold atoms, it is worth mentioning the possibility of trapping them in highly anisotropic optical lattices, thus realizing systems with different geometries and dimensions. In particular, strongly anisotropic lattices allow to realize systems in which the atoms are forced to live in one-dimensional (1D) lattices, i.e. on a chain~\cite{bloch2008,cazalilla2011}. 

From a theoretical point of view, physics in 1D is of particular interest not only because it may realize toy models for higher dimensional problems, but also because some physical effects, such as quantum ones, are much stronger and give rise to new phenomena and new states of matter~\cite{gogolin_book,giamarchi_book}. Analitically, one can use powerful low-energy field theories to describe fermionic, bosonic, and spin models in terms of collective bosonic degrees of freedom~\cite{gogolin_book,giamarchi_book}. Also, under suitable hypothesis, one can establish an equivalence between fermionic and spin degrees of freedom. 

Theoreticians have long been studying models by switching from a bosonic to a fermionic representation, from a fermion to a spin one, or viceversa. These mappings allow to see how new phases of matter may stabilize for certain values of the coefficients appearing in the Hamiltonian. This is the case, for example, of the emergence of magnetic ordered phases in fermionic and in bosonic models, a problem that was studied long ago in the seminal work by Mott. It was first demonstrated that  an insulating antiferromagnetic phase may arise in an apparently simple fermionic model such as the Hubbard one (see \cite{bala,essler} for reviews). The analysis has then been extended to more complicated systems, such as mixtures of fermionic species described by the two-species Hubbard model, where also more exotic phenomena may appear (singlet superconductivity, FFLO phase, super-counter-flow, etc.) (see Refs. \cite{bloch2008,cazalilla2011,feiguin2011} for a complete review). 

As for the bosonic case, it was first in Refs.\cite{giamarchi1988,fisher1989} that it was shown that the Bose-Hubbard model could admit a quantum phase transition from superfluid to an insulating magnetic Mott-like phase. Since then, much attention has been devoted to understand the features of this transition in arbitrary dimensions or in more complex models describing for example mixtures of bosonic species \cite{kuklov2003,altman2003,capogrosso,guglielmino,roscilde2012,hubener,hu,mathey,dalmonte2012,roscilde,isacsson}. 

In recent years, these theoretical predictions have become accessible to experimental checks in cold atoms experiments. Starting from the breakthrough demonstration of Mott-insulator to superfluid transition in a gas of Rb atoms \cite{greiner2002}, strongly correlated regimes have been systematically accessed in ultracold gas experiments, prominent examples being the realization of Tonks-Girardeau \cite{paredes2004,kinoshita2004} and super-Tonks gases \cite{haller2009}, the observation of 1D Mott \cite{stoeferle2004} and Berezinskii-Kosterlitz-Thouless transitions \cite{haller2010}, and various dynamical processes \cite{bloch2008,kinoshita2006}.

Even more intriguing is the possibility of investigating magnetism in such controllable setups, as recently shown in the case of the Ising model in  Ref.~\cite{simon2011}. In particular, two-species bosonic and fermionic gases represent ideal setups for the search of strongly correlated magnetic states of matter, since once the charge degrees of freedom are gapped in a Mott insulating region, the many-body dynamics is dominated by pseudo-spin degrees of freedom \cite{kuklov2003,altman2003,guglielmino}, as in the case of the aforementioned Hubbard model~\cite{essler}. 

In this work, motivated by impressive experimental achievements in tuning and controlling low-dimensional heteronuclear bosonic mixtures~\cite{catani2008,catani2009}, we present a systematic and quantitative investigation for the realization of magnetic phases in a feasible setup of a Bose-Bose mixture of cold atoms trapped in one-dimensional optical lattices. In Sec. \ref{sec:modelham}, we will describe how to model the system via a two-species Bose-Hubbard Hamiltonian,  whose effective  parameters are determined by microscopic properties, such as the depth of the  lattice potential and the Feshbach resonance scattering length. We will discuss this relationship in details for Rb-K mixtures which, thanks to their high tunability \cite{catani2008}, allow to span a wide range for both the hopping and the interaction coefficients.  

In Sect. 3, we will discuss the theoretical background to study the emergence of magnetic phases in such a system. To describe the physics deep inside the insulating Mott regions, where the single site density is fixed to integer values, we will study the two species Bose-Hubbard Hamiltonian in the strong coupling regime via a perturbative scheme which is a generalization of the Schrieffer-Wolff transformation~\cite{schr}. At integer fillings, the so obtained effective Hamiltonian can then be mapped onto a spin model, which turns out to be the spin-1/2 $XXZ$-chain for filling one particle per site and the $\lambda-D$ spin-1 Hamiltonian for filling two. Both these models are paradigmatic for the description of many body systems that display quantum phase transitions~\cite{giamarchi_book,chen2003,kennedy}. As it is well known, for strong anisotropies, the former one admits 
a ferromagnetic and an antiferromagnetic regime. The latter has a very interesting zero-temperature phase diagram, with different critical and massive phases, which include ferromagnetic, Ising-like as well as spin-0 ground states. In particular, it displays a Haldane phase, which is determined by the breaking of a non-local symmetry and described by non-local order parameters, the so-called string-order ones. In both cases, we will address how feasible is to reach these regimes from an experimental point of view.

In Sec. \ref{sez4}, we will study the Bose-Hubbard Hamiltonian in the spin-$1/2$ magnetic sector by means of numerical simulations based on the density-matrix-renormalization-group (DMRG) algorithm~\cite{white1992,schollwoeck2005}. 
Our calculations allow to determine a quantitative phase diagram for the experimentally relevant case of Rb-K mixtures. On the basis of this analysis, we will be able to predict the emergence of the various massive phases as the parameters change in an experimentally feasible range, and estimate the critical temperatures needed to achieve such phases in experiments. 

Finally, in Sec. 5, we draw our conclusions and discuss possible developments.

\section{Model Hamiltonian and experimental regimes}
\label{sec:modelham}
Two-species Bose mixtures strongly confined by optical lattice potentials in two-directions and
subject to an additional periodic potential along the third direction are well described by the 
following microscopic Hamiltonian:
\begin{equation}
\begin{split}
H=&\int dx \sum_{\alpha=1,2}\psi^\dagger_{\alpha}(x)\left[-\frac{\hbar^2\partial_x^2}{2m_\alpha} +V(x)+\mu_\alpha(x)\right]\psi_\alpha(x)+\\
&+\int dx dx' \sum_\alpha \;\rho_\alpha(x)\rho_\alpha(x')\mathcal{U}_{\alpha}\delta(x-x')+\\
&+\int dx dx'\;\rho_1(x)\rho_2(x')\mathcal{U}_{12}\delta(x-x')
\end{split}
\end{equation}
Here, $\psi^\dagger_\alpha, \psi_\alpha$ are bosonic creation/anihilation operators, 
$m_\alpha$ are particle masses, $V(x)=V\sin^2(2\pi x/\lambda)$ is the optical lattice potential along the wire of depth $V$ and wavelength $\lambda$, and $\mu(x)$
the confining potential along the $x$-direction.
The intra-species interaction strengths $\mathcal{U}_{\alpha}=2\hbar\omega_{\perp,\alpha} a_{1D,\alpha}$ are related to the intraspecies 1D scattering length $a_{1D,\alpha}$, with $\omega_{\perp,\alpha}=\frac{\hbar\pi}{2m_\alpha d^2}\sqrt{V_\perp/E_{r,\alpha}}$ 
being the transverse confinement frequency depending on both the optical lattice
spacing $d=\lambda/2$ and the intensity of the transverse field $V_\perp$ in units
of the recoil energy $E_{r,\alpha}=\frac{h^2}{8m_\alpha d^2}$. Finally, the inter-species
interaction $\mathcal{U}_{12}$ is usually tunable via Feshbach resonances~\cite{chin2010}, and 
is proportional to the inter-species scattering length $a_{1D, 12}$. In the limit of sufficiently deep lattice potentials along the wire direction, $V/E_{r,\sigma}\gtrsim 6$, the system is well modeled by an effective Bose Hubbard Hamiltonian:
\begin{equation*}
\begin{split}
H_{BB}=&-\sum_\sigma t_\sigma \sum_i (a_{i,\sigma}^{\dagger}a_{i+1, \sigma} + \text{h.c.}) +U_{ab}\sum_in_{i,a}n_{i,b}\\
&+\sum_{\sigma}\frac{U_\sigma}{2}\sum_i n_{i,\sigma}(n_{i,\sigma}-1)+\sum_\sigma \mu_{\sigma}\sum_i n_i (i-L/2)^2
\end{split}
\end{equation*}
where the first term represents tunneling processes of both species $\sigma=1,2$, the second and third 
are the inter- and intra-species interaction, and the last one is a species-dependent confining potential.
The operators $a^\dagger_{j,\sigma}, a_{j,\sigma}$ satisfy bosonic commutation relations, and 
$n_{i,\sigma}=a^\dagger_{i,\sigma}a_{i,\sigma}$. The effective Hubbard parameters can be 
derived from microscopic  quantities by expanding the single particle wave-functions in the 
Wannier basis, thus obtaining the following relations~\cite{bloch2008}:
\begin{gather}\label{t_micro}
t_\sigma=\frac{4E_{r,\sigma}}{\sqrt{\pi}}\left(\frac{V}{E_{r,\sigma}}\right)^{3/4}\exp\left[-2 \left(\frac{V}{E_{r,\sigma}}\right)^{1/2}\right] \\
U_\sigma=\frac{\sqrt{2\pi}\mathcal{U}_\sigma}{\lambda}\left(\frac{V}{E_{r,\sigma}}\right)^{1/4}\label{U_micro}
\end{gather}
Once a specific experimental setup is considered, the relevant Hamiltonian parameters 
can be directly shaped by means of external fields as follows: the hopping rate ratio
is tuned by considering, e.g., species dependent lattices, or, in case of species
independent one, by just changing the lattice depth (which affects more drastically
heavier particles). Interactions can be then tuned by Feshbach resonances; in 
particular, leaving intra-species interactions unmodified, it is possible to span a broad range
of interaction strengths by means of an external magnetic field, or, eventually, employing
confinement induced resonances.

\subsection{Experimental parameters for Rb-K mixtures}
As a case study, we will focus from here onwards on $^{87}$Rb-$^{41}$K mixtures.
This choice is justified, from the one hand, by the notable experimental achievements 
already demonstrated with these mixtures such as, e.g., full tunability of interactions~\cite{catani2008},
and, on the other hand, by the relative large mass-imbalance, which allows to span 
parameters regimes ranging from intermediate to large hopping ratios. Here, 
we consider optical lattices with wavelength $\lambda/2=532$nm (as reported 
in~\cite{catani2008}), 
and define dimensionless optical lattice depths $s=V/E_{r,K}$ and $s_\perp=V_\perp/E_{r,K}$
along the $x$ and $y-z$ directions respectively.

\begin{figure}
\begin{center}
\includegraphics[scale=0.5]{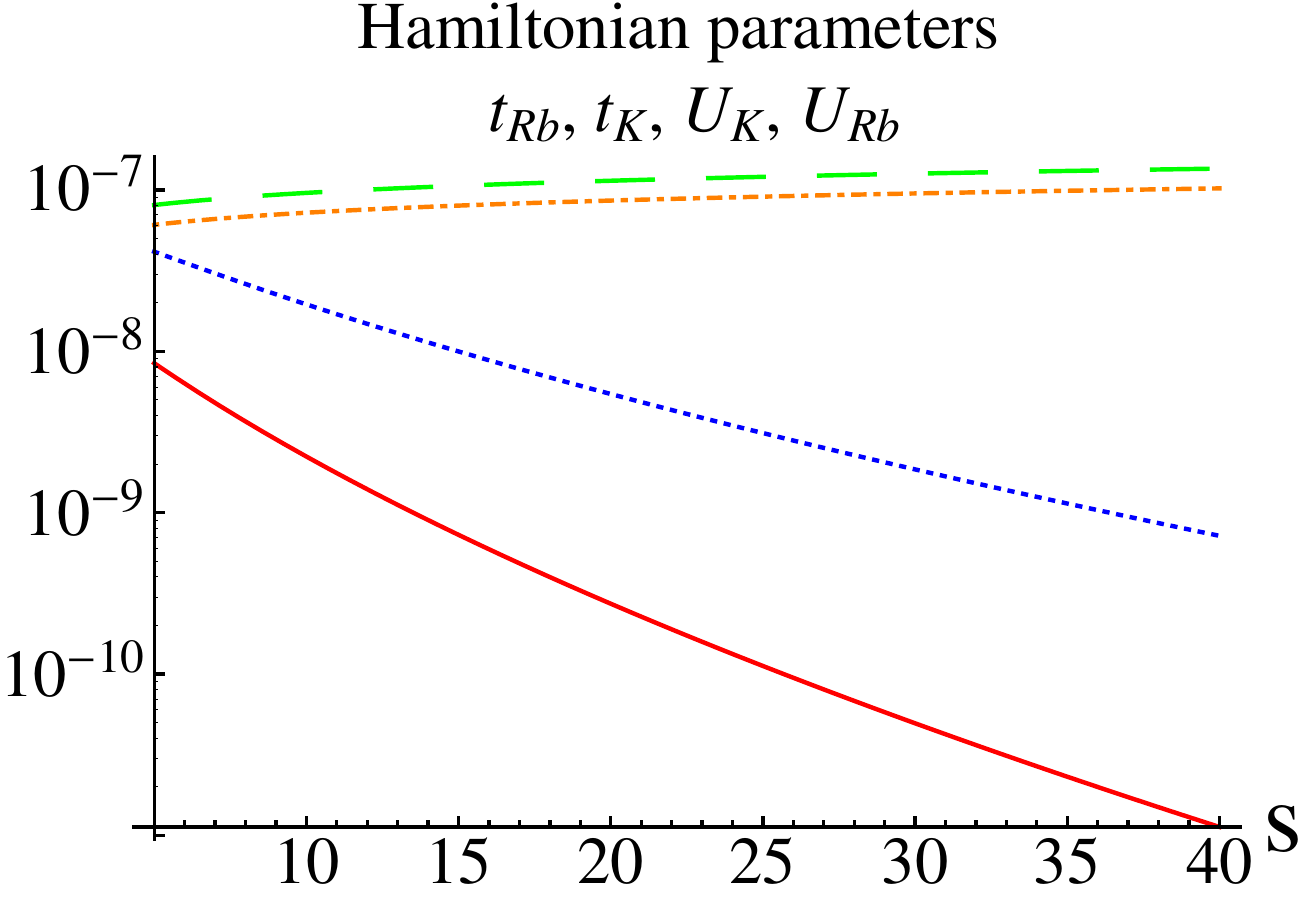}%

\includegraphics[scale=0.45]{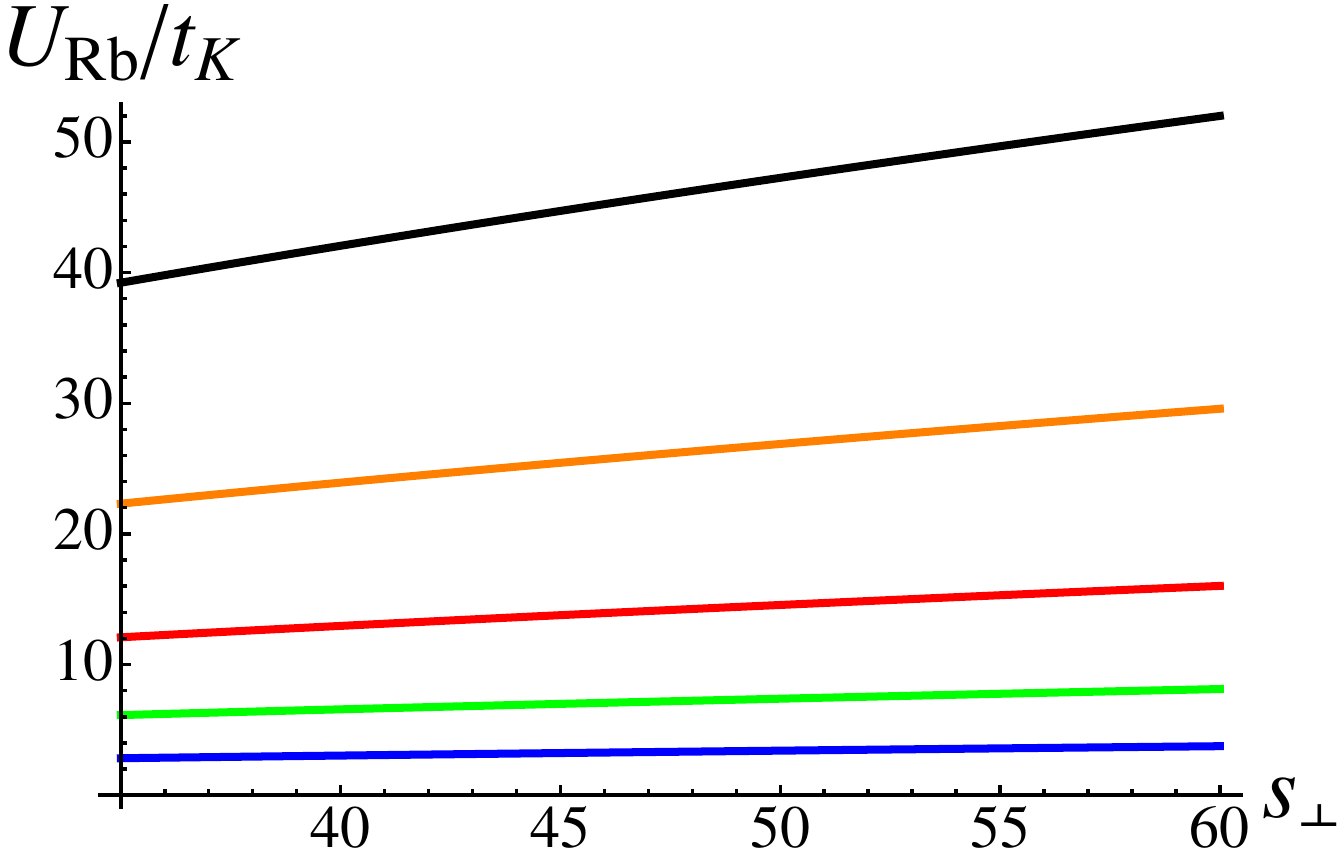}
\end{center}
\caption{Typical Hubbard parameters for $^{87}$Rb-$^{41}$K mixtures. Upper panel:
Strength of the different Hamiltonian parameters in K units as a function of the 
lattice depth along the wire: $t_{Rb}$ (thin, red), $t_{K}$ (blue, dotted), 
$U_{K}$ (dot-dashed, orange), and $U_{Rb}$ (dashed, green). The tunneling rates 
decrease exponentially when increasing $s$.
Lower panel: intra-species interaction scaling a function of the orthogonal lattice depth 
$s_\perp$ for different values of $s=10, 15, 20, 25$ from bottom to top.} 
\label{fig:mix_parameters}
\end{figure}

Typical experimental parameters obtained from Eq.~\ref{t_micro}, \ref{U_micro} are presented 
in Fig.~\ref{fig:mix_parameters}. The tunneling rates are strongly
suppressed as a function of $s$, as expected, and in general $U_{Rb}>U_K$.
Moreover, in the lower panel, it is shown how inter particle interactions can be 
increased by considering stronger confinement along $y-z$ directions. The regimes we will be mainly
interested in, that is, the ones where magnetic properties may emerge, 
have to be characterized by sufficiently strong intra-species interactions and not too
small hopping rates, so that robustness with respect to temperature is stronger.
As such, intermediate values of $s\simeq 10-25$ and possibly large values of 
$s_\perp$ will be our focus from here on.

\section{Perturbation theory at integer filling}
\label{sec:3}
In this section, we analyse the magnetic phases deep inside the Mott regions at 
integer filling, $n_{Rb}+n_{K}\in \mathbb{N}, n_K=n_{Rb}$, and discuss the qualitative 
phase diagrams realistically achievable with Rb-K mixtures. Our treatment flows along
the lines of Ref. \cite{altman2003}, where the possible magnetic phases of ultra cold atomic mixtures have also
been discussed. 

In the strong coupling limit ($t_1, t_2 \ll U_{RbK},$ $U_1,U_2$, where we have defined 1=K, 2=Rb for compactness), processes in which single particle tunneling changes the total on-site populations require high energy and thus the low energy Hilbert subspace $\Lambda$ contains only states with a fixed integer occupation number on every site. To derive an effective Hamiltonian acting in such subspace we use a generalization of the Schrieffer-Wolff transformation \cite{schr,barbiero2010}. 

This kind of technique may be used whenever one has to deal with a Hamiltonian of the type: $H = H_0 + V$, where the unperturbed Hamiltonian $H_0$ may be diagonalized  within a Hilbert subspace $\Lambda$ \cite{bala}.  In our case, $V$ contains the hopping terms, whereas two-body interactions determine $H_0$. Here, $\Lambda$ is the subspace with fixed occupation number for every site $n_i \equiv \nu$. 
By denoting with $P, P^\perp$ the orthogonal projectors on the subspaces $\Lambda, \Lambda^\perp$ respectively, one may write the Hamiltonian as the sum of a diagonal and an off-diagonal term: $H= {\cal H}_0 + {\cal H}_1$ with
\begin{eqnarray}
&\mathcal{H}_0 = PH_0P + P^\perp H_0P^\perp \\
&\mathcal{H}_1 = PVP^\perp +P^\perp VP +P^\perp VP^\perp .
\end{eqnarray}
Then one looks for a unitary transformation $U = \exp (i\epsilon S)$, with $S^\dagger = S$, that diagonalizes the total Hamiltonian. The operator $S$ may be found by imposing $[{\cal H}_0, S] + i {\cal H}_1 = 0$,
meaning that, up to the first non-trivial correction in $\epsilon$, the diagonal form of $H$ is finally given by:
\begin{equation}
H_{\text{eff}} = {\cal H}_0 + i [S, {\cal H}_1].
\end{equation}

\subsection{Spin-1/2 chains for $\nu=1$}
Deep inside the first Mott lobe, when $\nu=1$, the Hilbert subspace $\Lambda$ is the tensor product of single-site Hilbert spaces which are two-dimensional and generated by the two states:
\begin{equation}
\lvert\uparrow_i \rangle \equiv a^\dagger_{i,1} \lvert\mathrm{vac}\rangle  \;\; , \;\; \lvert\downarrow_i \rangle \equiv a^\dagger_{i,2}\lvert\mathrm{vac}\rangle \; .
\end{equation}
On this space one can define the spin-1/2 operators:
\begin{gather}
 S^x_i =  \left(\lvert\uparrow_i \rangle \langle\downarrow_i\rvert + \lvert\downarrow_i \rangle \langle\uparrow_i\rvert \right)/2\\
S^y_i =\left( \lvert\uparrow_i \rangle \langle\downarrow_i \rvert - \lvert\downarrow_i \rangle \langle\uparrow_i\rvert\right)/2i\\
S^z_i = \left( \lvert\uparrow_i \rangle \langle\uparrow_i \rvert- \lvert\downarrow_i \rangle \langle\downarrow_i\rvert\right)/2
\end{gather}
which allow to rewrite the Schrieffer-Wolff transformed Hamiltonian as:
\begin{eqnarray}
H & = & \sum_i \left[  J_z S^z_i S^z_{i+1}  - J_\perp \left( S^x_i S^x_{i+1} +S^y_i S^y_{i+1} \right) \right. \\
& - & \left. h\left(S^z_i +S^z_{i+1}\right) -  \lambda \mathbb I   \right] \nonumber
\end{eqnarray}
where
\begin{equation}\label{Jperp}
 J_z =2  \frac{t_1^2+t_2^2}{U_{RbK}} -  \frac{4t_1^2}{U_1} -  \frac{4t_2^2}{U_2}, \; J_\perp =  \frac{4t_1 t_2}{U_{RbK}},   
 \end{equation}
 
\begin{equation}
h =  \frac{2 t_1^2}{U_1} -  \frac{2 t_2^2}{U_2} , \; \lambda = \frac{t_1^2}{U_1} + \frac{t_2^2}{U_2} +  \frac{t_1^2+t_2^2}{2U_{RbK}}.\label{Jz}  
\end{equation}
Since in any canonical ensemble, $h$ is effectively fixed to~$0$ (the effects of the trapping 
potential will be discussed below), we see that, up to a constant factor, the Hamiltonian represents a one-dimensional spin-1/2 XXZ model, as first found in \cite{altman2003}. Interestingly, this basic model displays already a rich phase diagram, characterized by phase transitions from a critical region for $ -1 \leq \Delta \equiv J_z/J_\perp \leq 1$ to massive phases, either ferromagnetic (FP) for $\Delta <-1$ or anti-ferromagnetic (AFP) for $\Delta >1$. In the bosonic language, the former is a super-counter-flow (SCF) superfluid phase, which can be interpreted as a superfluid of particle-hole bound states, as easily seen by performing a particle-hole transformation in one of the species~\cite{kuklov2003}.

\begin{figure}[t]\centering
\includegraphics[width=0.6\textwidth]{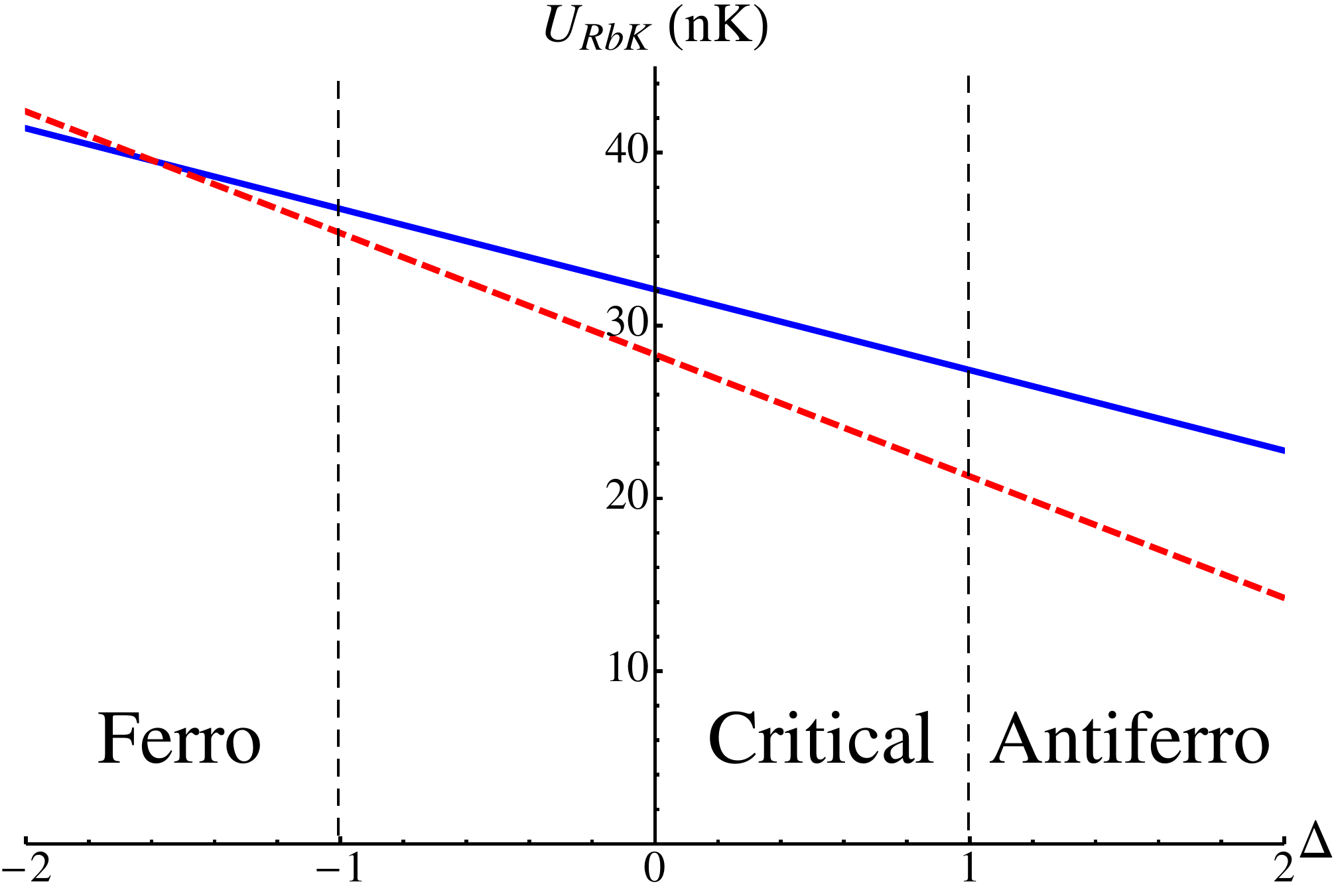}
\caption{$\nu=1, S=1/2$ case. Evolution of $\Delta=J_z/J_\perp$ with respect to the interspecies interaction $U_{RbK}$ for $s=9, s_\perp=65$ (red, dashed line) and  $s=15, s_\perp=65$ (blue, solid line). The system can be driven from a ferromagnetic to a critical and, eventually, to an anti-ferromagnetic phase, by decreasing $U_{RbK}$.}
\label{par_s12}
\end{figure}

As shown in Fig. \ref{par_s12}, the entire phase diagram can be scanned by just changing
the interspecies interaction via Feshbach resonances. For very large values of $U_{RbK}$,
the system displays ferromagnetic behavior, corresponding in the atomic language to real-space 
phase separation; on the other hand, weak repulsion will indeed favor short-range
antiferromagnetic interactions, due to the positive value of the diagonal interaction $J_z$,
as can be noticed from Eq.~\eqref{Jz}.

\subsection{Spin-1 chains for $\nu=2$ and the Haldane phase}

In this section, we analyse the magnetic phases deep inside the Mott region with $\nu=2$. In this case the low-energy Hilbert subspace is the tensor product of single-site Hilbert spaces which are generated by three states:
\begin{gather}
\lvert -1_i\rangle = \frac{(a^\dagger_{i,1})^2}{\sqrt{2}} \lvert\mathrm{vac}\rangle, \;  \lvert +1_i\rangle = \frac{(a^\dagger_{i,2})^2}{\sqrt{2}} \lvert \mathrm{vac}\rangle, \nonumber \\  \lvert 0_i\rangle = a^\dagger_{i,1} a^\dagger_{i,2} \lvert\mathrm{vac}\rangle 
\end{gather}
which represent a multiplet for the spin-1 operators \cite{altman2003}:
\begin{eqnarray}
& S^+_i =  \sqrt{2} \left( \lvert +1_i \rangle \langle  0_i \rvert + \lvert  0_i \rangle \langle -1_i \rvert \right)\\
& S^-_i =  \sqrt{2} \left( \lvert -1_i \rangle \langle  0_i \rvert + \lvert  0_i \rangle \langle +1_i \rvert \right)\\
& S^z_i =                  \lvert +1_i \rangle \langle +1_i \rvert - \lvert -1_i \rangle \langle -1_i \rvert 
\end{eqnarray}
By means of the Schrieffer-Wolff transformation, one obtains an effective spin-1 Hamiltonian which, neglecting constant and effective magnetic field terms, is given by:
\begin{equation}\label{eqn:HamiltonianDoubleFilling}
\begin{split}
H_\mathrm{eff}=& \sum_{i}\eta (S_i^xS_{i+1}^x+S_i^yS_{i+1}^y) + \lambda S_i^z S_{i+1}^z + D(S_i^z)^2  \\ & +\delta (S_i^z S_{i+1}^z)^2 +\zeta (S_i^z(S_{i+1}^z)^2+(S_i^z)^2S_{i+1}^z)  \\
& + \chi(\Gamma_i^{xz}\Gamma_{i+1}^{xz}+\Gamma_i^{yz}\Gamma_{i+1}^{yz})
\end{split}
\end{equation}
where $\Gamma_i^{\alpha\beta}=S^\alpha_iS^\beta_i+S^\beta_i S^\alpha_i $. In terms of
the original parameters, the couplings are given by:
{\small
\begin{align*}
\eta = &\frac{1}{2} \left(  \frac{t_1 t_2}{U_1 - 2U_{RbK}} + \frac{t_1 t_2}{U_2 - 2U_{RbK}}  -  \frac{t_1 t_2}{U_1 } -  \frac{t_1 t_2}{U_2}  \right)-  \frac{2t_1 t_2}{U_{RbK}} \\
\lambda = & - \frac{3t_1^2}{U_1} - \frac{3t_2^2}{U_2} +  \frac{t_1^2}{2U_{RbK}-U_1} +  \frac{t_2^2}{2U_{RbK}-U_2} \\
D = & \frac{U_1+U_2 -2U_{RbK}}{2} + \frac{t_1^2}{2} \left( \frac{3}{U_{RbK}-2U_1} +\frac{8}{U_1} - \frac{5}{U_{RbK}} \right)\\
&+ \frac{t_2^2}{2} \left( \frac{3}{U_{RbK}-2U_2} +\frac{8}{U_2} - \frac{5}{U_{RbK}}  \right)\\
\delta = & t_1^2 \left( \frac{3}{U_{RbK}-2U_1} +  \frac{1}{U_1-2U_{RbK}}- \frac{7}{U_1} + \frac{5}{U_{RbK}}  \right) \\
& +t_2^2 \left( \frac{3}{U_{RbK}-2U_2} +  \frac{1}{U_2-2U_{RbK}}- \frac{7}{U_2} + \frac{5}{U_{RbK}}  \right) \\
\begin{split}\zeta = &\frac{t_1^2}{2} \left(  - \frac{3}{U_{RbK}-U_1} - \frac{6}{U_1} + \frac{3}{U_{RbK}}   \right) \\ & + \frac{t_2^2}{2} \left(  \frac{3}{U_{RbK}-U_2} + \frac{6}{U_2} -\frac{3}{U_{RbK}}   \right) \end{split} \\
\chi =&\frac{t_1 t_2}{2U_1 } +  \frac{t_1 t_2}{2U_2} - \frac{t_1 t_2}{2U_1 - 4U_{RbK}} - \frac{t_1 t_2}{2U_2 - 4U_{RbK}}  -  \frac{2t_1 t_2}{U_{RbK}}.
\end{align*}}%

\begin{figure}\centering
\includegraphics[width=0.6\textwidth]{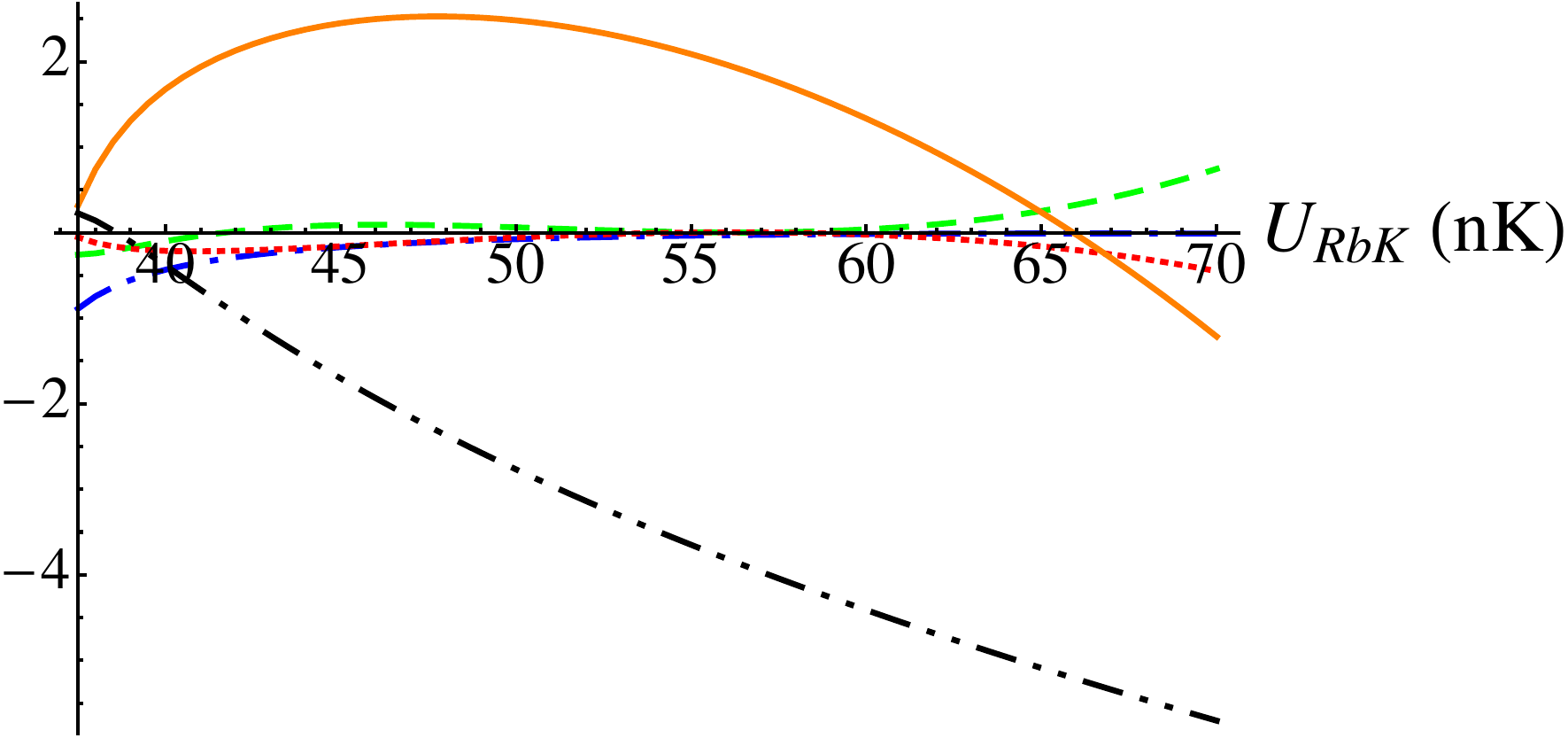}
\caption{$\nu=2, S=1$ case. The parameters of the effective Hamiltonian \eqref{eqn:HamiltonianDoubleFilling} as a function of the interspecies interaction $U_{RbK}$ for $s=9, s_\perp=65$ and assuming $|\eta|=1$: $\lambda$ (black, dashed-dotted-dotted), $D$ (orange, solid), $\delta$ (green, dashed), $\zeta$ (red, dotted) and $\chi$ (blue, dashed-dotted). Here, the hopping parameter is $t_K\simeq22$nK.}
\label{fig:AllParameters}
\end{figure}

In Fig.~\ref{fig:AllParameters}, we show how the different parameters (measured in units of \(|\eta|\))
depend on the interspecies interaction $U_{RbK}$ for different values of \(s\) and \(s_\perp\). 
In a relatively large parameter ($U_{RbK}$) region, \(\delta, \zeta, \chi\) are negligible with respect to \(\lambda\) and \(D\), so the Hamiltonian reduces to that of the so-called \(\lambda-D\) model \cite{kennedy,schulz1986}:
\begin{equation}\label{eqn:HamiltonianXXZSpinUno}
H_{\lambda-D}=\sum_i\eta(S_i^x S_{i+1}^x+ S_i^y S_{i+1}^y) + \lambda S_i^z S_{i+1}^z + D (S_i^z)^2. 
\end{equation}
For example, if \(s=15, s_\perp=65\), this interval is given by \(43 \mathrm{nK} \lesssim U_{{RbK}} \lesssim 70 \mathrm{nK}\), while if \(s=9, s_\perp=65\) it is given by \(36 \mathrm{nK} \lesssim U_{{RbK}} \lesssim 60 \mathrm{nK}\).

The ground state phase diagram of the \(\lambda-D\) model is even richer than the $S=1/2$ Heisenberg model one. Some qualitative considerations can be easily made when the first term in eq. (\ref{eqn:HamiltonianXXZSpinUno}) can be neglected: if \(\lambda\) is larger than \(D\), the ground state reduces to a ferromagnetic state (if \(\lambda<0\)) or antiferromagnetic state (if \(\lambda>0\)); if \(D \gg |\lambda| \) (large--\(D\) phase), spins in all sites prefer to have spin zero. For values of $|D|$ which are comparable with $|\lambda|$, instead, the first term in eq. (\ref{eqn:HamiltonianXXZSpinUno})  is important and new phases emerge, as shown numerically in \cite{chen2003}, 
separated by critical lines with different universality classes \cite{esposti}. More specifically, if $\lambda$ is negative, the system is critical, belonging to the XX universality class, while for $\lambda$ positive a new massive phase arises, typical of isotropic spin-1 chain: the Haldane phase \cite{haldane}.

To characterize all massive phases, one may exploit the symmetries of the Hamiltonian. Firstly, it has an explicit $Z_2$-symmetry to describe which one can consider the \textit{ferromagnetic order parameter} and the \textit{N\'eel order parameter} (\(\langle\cdot\rangle\) means expectation value in the ground state):
\begin{gather}
\mathcal{O}^\alpha_{\mathrm{FM}} = \lim_{|i-j|\rightarrow \infty}\langle{S^\alpha_iS^\alpha_j}\rangle; \\
\mathcal{O}^\alpha_{\mathrm{N\acute{e}el}} = \lim_{|i-j|\rightarrow \infty}\langle(-1)^{|i-j|}S^\alpha_i S^\alpha_j \rangle.
\end{gather}
As shown in \cite{kennedy}, the $\lambda-D$-Hamiltonian has also a hidden $Z_2\otimes Z_2$-symmetry which corresponds to a set of non-local order parameters, the so called \textit{string order parameters} \cite{dennijs1989}:
\begin{equation}
\mathcal{O}^\alpha_{\mathrm{string}} = \lim_{|i-j|\rightarrow \infty}\langle{-S^\alpha_i e^{i \pi \sum_{l=i+1}^{j-1}S^\alpha_l} S^\alpha_j}\rangle,
\end{equation}
where $\alpha=x,y,z$. The magnetic phase diagram can then be schematically represented as follows:
{\it i)} in the ferromagnetic phase, only \(\mathcal{O}^z_{\mathrm{FM}}\) is non zero, while all the other parameters are zero; {\it ii)} in the N\'eel phase, \(\mathcal{O}^\alpha_{\mathrm{string}}=\mathcal{O}^\alpha_{\mathrm{N\acute{e}el}}=0\) for \(\alpha=x,y\), \(\mathcal{O}^\beta_{\mathrm{FM}}=0\) for all \(\beta\), while  \(\mathcal{O}^z_{\mathrm{string}}\) and \(\mathcal{O}^z_{\mathrm{N\acute{e}el}}\) are non zero; {\it iii)} in the large--\(D\) phase, all the order parameters are zero. Finally, {\it iv)} in the Haldane phase, all three of \(\mathcal{O}^\alpha_{\mathrm{string}}\) are non zero, while all of \(\mathcal{O}^\alpha_{\mathrm{N\acute{e}el}}\) and of \(\mathcal{O}^\alpha_{\mathrm{FM}}\) are zero, so that this phase can be seen as a dilute antiferromagnet, 
where antiferromagnetic correlations survive despite loosing the associated positional order present in 
the N\'eel phase.

\begin{figure}\centering
\includegraphics[width=0.6\textwidth]{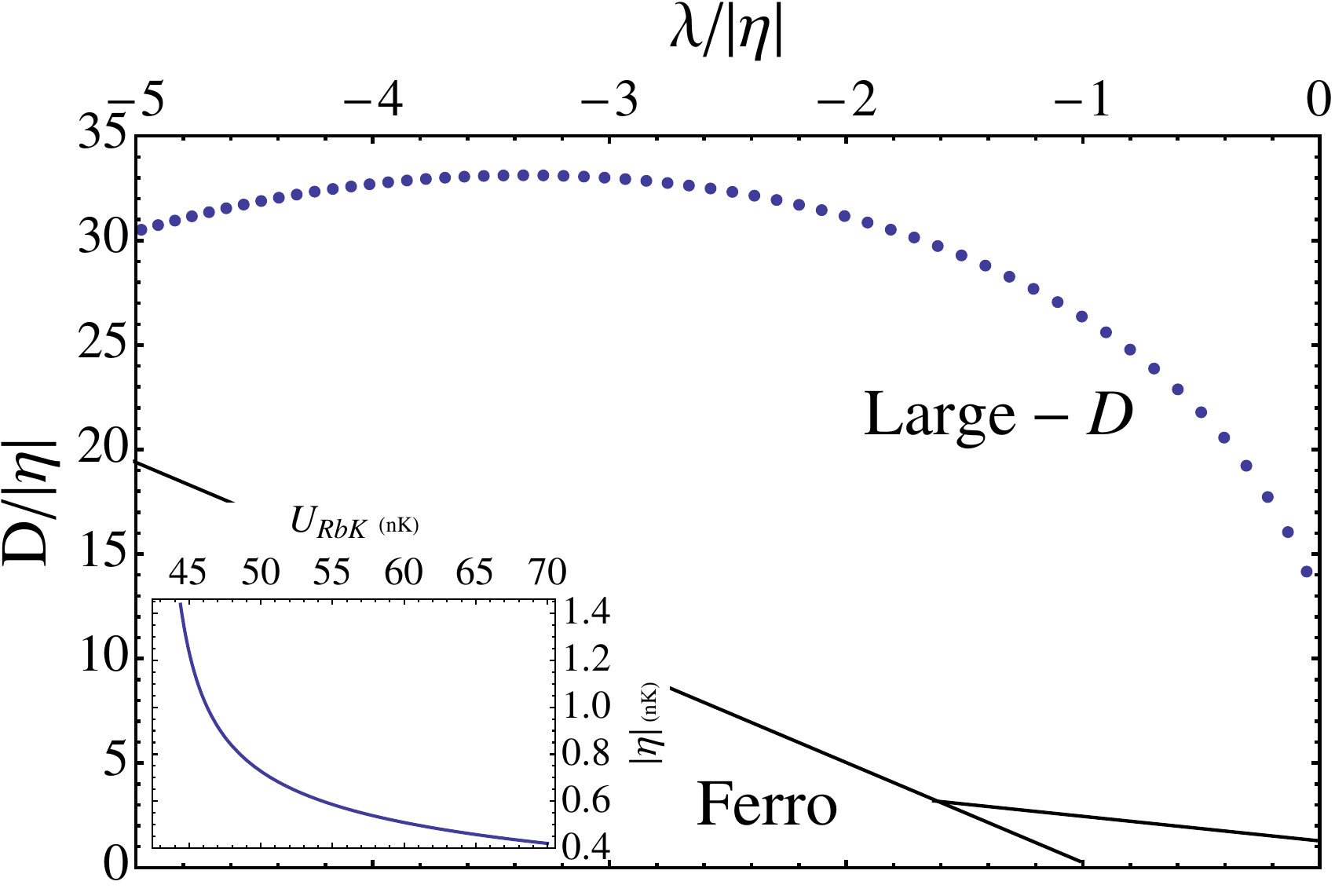}%

\includegraphics[width=0.6\textwidth]{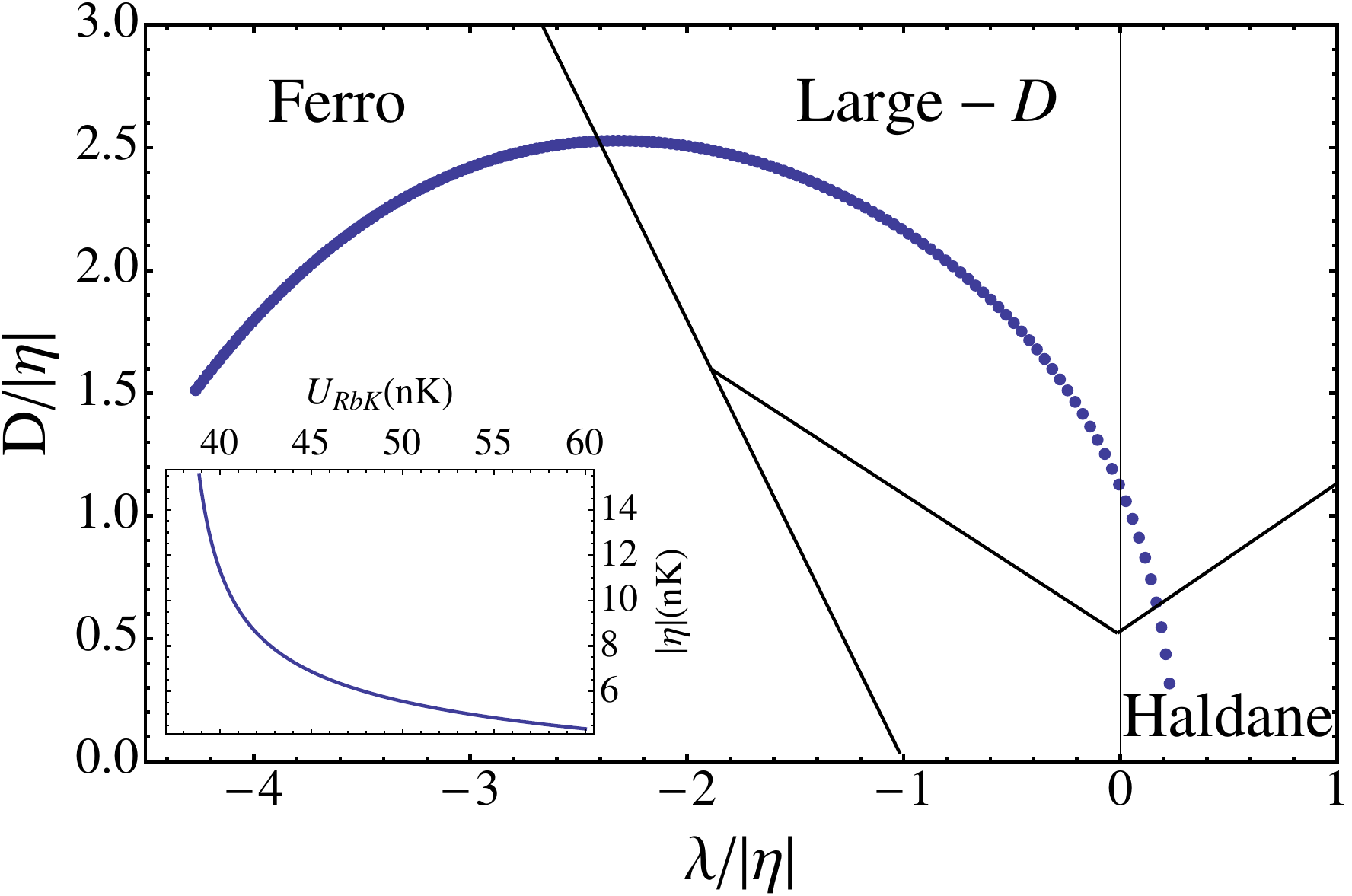}
\caption{Phase diagram of Hamiltonian \eqref{eqn:HamiltonianXXZSpinUno}  for \(s=15, s_\perp=65\) (upper) and \(s=9, s_\perp=65\) (lower panel). The dots represent the ground state of \eqref{eqn:HamiltonianXXZSpinUno} by varying the interspecies interaction \(U_{RbK}\) (43nK\(<U_{RbK}<\)70nK, lower 38nK\(<U_{RbK}<\)60nK) . When \(U_{RbK}\)  increases the ratio $\lambda/|\eta|$ becomes more negative in both cases. In both insets the parameter \(\eta\), fixing the energy scale, is plotted versus \(U_{RbK}\).} \label{fig:PhaseDiagramLambdaD}
\end{figure}

From the formulas given above, one can estimate the values of $\lambda$ and $D$ and thus find that systems with \(s\) large (\(s\ge 15\))  will be mainly in the large--\(D\) phase, as it is shown in \figurename~\ref{fig:PhaseDiagramLambdaD} (upper panel) for the case with \(s=15, s_\perp=65\) and \(U_{RbK}\) in the interval \(\sim 43/70 \mathrm{nK}\). When the value of \(s\) decreases, it is possible to go beyond the large--\(D\) phase and, in principle, reach the Haldane phase, as one can see from \figurename~\ref{fig:PhaseDiagramLambdaD} (lower panel) for \(s=9, s_\perp=65\) and in a small interval of \(U_{RbK}\).

Nevertheless, we notice that the range of \(U_{RbK}\) in which the model reduces to the \(\lambda-D\) one is not wide. Also, the Haldane phase region can be reached only for small values of $s$, where other terms in the effective Hamiltonian may compete with $\lambda,D$ and affect (or even destroy) topological order. Clearly, both these constraints 
represent a serious challenge for experimental realizations of the Haldane phase, as a clear observation would require both fine tuning of the parameters and very low temperatures. Finally, since the optical lattice depth in these cases would be of order $s\simeq9$, it would be preferable to check second order perturbation theory results against more accurate methods such as DMRG or Quantum Monte Carlo simulations.

\section{Numerical results}
\label{sez4}
In order to complement perturbation theory predictions, we will present
here numerical results for the realistic Hubbard parameters discussed
in Sec.~\ref{sec:modelham} and equal filling for both species $n_{Rb}=n_K=1/2$. 

Here, we are interested in 
{\it i)} determine a quantitative phase diagram spanning a broad
range of interaction parameters, and {\it ii)} determine the optimal
parameter regime in terms of resilience to thermal fluctuations, 
that is, identify the interaction strength around which the smallest
between the charge and spin gaps,  $\Delta_c, \Delta_s$ is larger in absolute (that is,
when expressed in nK units) value. This last observation is 
particularly interesting in view of possible experimental investigations, 
as it provides a quantitative guide to understand the critical
temperatures needed to observe magnetic phases.
Moreover, such information is not accessible within perturbation 
theory, as the largest gap region usually lies outside of its 
applicability regime, as shown below.

All simulations have been carried employing the DMRG algorithm, 
a state-of-the-art method to tackle ground state properties of 1D systems. 
In order to determine the quantitative phase diagram of the system
and, at the same time, provide relevant energy scales for both charge
and spin degrees of freedom, we have investigated the
behavior of the charge:
\begin{equation}
\Delta_{c} (L) = \frac{1}{2}\left[E_L(L/2+ 1, L/2+1) + E_L(L/2- 1, L/2 - 1)-2 E_L (L/2, L/2)\right],
\end{equation}
and spin
\begin{equation}
\Delta_{s} (L) =\frac{1}{2}\left[E_L(L/2- 1, L/2+1) + E_L(L/2+ 1, L/2 - 1)-2 E_L (L/2, L/2)\right], 
\end{equation}
gap, where $E_L(N_1, N_2)$ is the ground state energy at size $L$ in the particle
number sector $(N_1,N_2)$. Far from the ferromagnetic regime
$U_{RbK}>U_{Rb}, U_K$, the extrapolated value of the two gaps
$\Delta_\alpha=\lim_{L\rightarrow\infty}\Delta_\alpha(L)$ is sufficient
to determine the magnetic phase diagram of the system 
according to Table ~\ref{tab:gap}. In particular, the N\'eel phase
is fully gapped, while the super-counter-flow phase, correspondent
to the critical phase of the XXZ model, has a finite $\Delta_c$ but
gapless spin excitations.

\begin{table}  
\centering
\begin{tabular}{||p{2cm}||*{3}{c|}|}
\hline                    
\bfseries        \centering{Phase}  & $\Delta_{\mathrm{C}}$ & $\Delta_{\mathrm{S}}$  \\
\hline
\hline
\bfseries       \centering{$\mathrm{AFM}$}       &    $\neq0$     &   $\neq0$   \\
\hline
\bfseries      \centering{$\mathrm{SCF}$}        &    $\neq0$      & 0    \\
\hline  
\bfseries      \centering{$\mathrm{SF}_{a}+\mathrm{SF}_{b}$}       &     0     &   0 \\
\hline
\end{tabular}
\caption{Strong coupling phases and relation with the charge and spin gaps.\label{tab:gap}}
\end{table}     

\begin{figure}
\begin{center}
\includegraphics[scale=0.28]{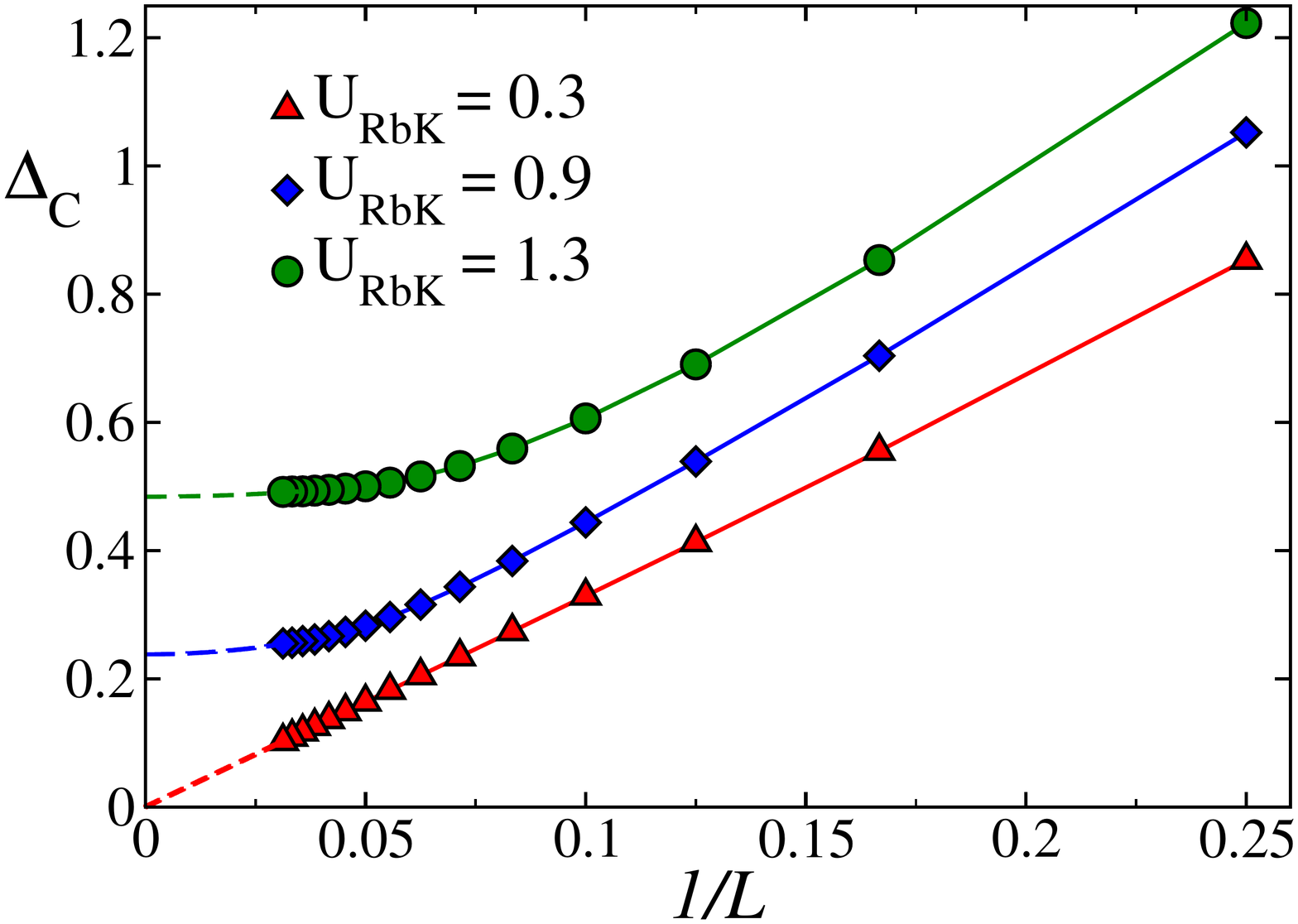}
\includegraphics[scale=0.28]{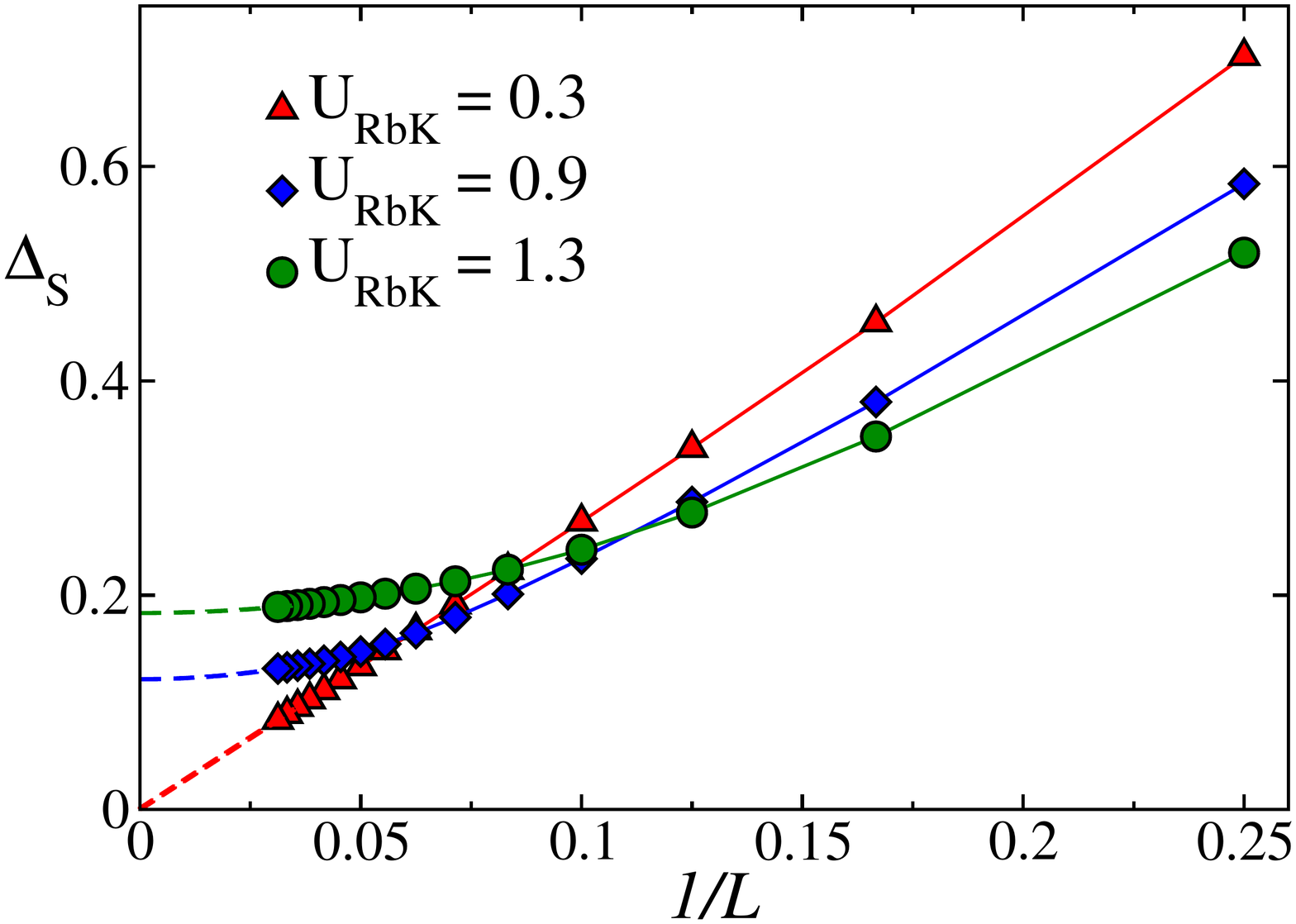}
\end{center}
\caption{Finite-size scaling analysis for both charge (upper panel) and spin 
(lower panel) gaps; here, $s=26, s_\perp=70$, and 
inter particle interactions are varied. In both panels, lines are linear ($U_{RbK}=0.3$) and algebraic
(others) best fits.} 
\label{fig:gaps}
\end{figure}

In order to suppress boundary effects, we have employed 
periodic boundary conditions (PBCs); despite being computationally more
requiring and limited to relatively small system sizes (here, up 
to $L=32$), they assure much better scaling properties since 
translational symmetry is not explicitly broken, as reflected in 
the accurate finite-size scaling procedure (illustrated
in Fig.~\ref{fig:gaps}). High accuracy
in ground state and excited state energies (with discarded weights
usually of order $5\times 10^{-6}$ or smaller)
have been achieved by employing up to 1024 states per
block and 5 finite-size sweeps, and further enhanced by applying an
iterative procedure which used the final finite-size step at size $L-2$ as the 
initial one at size $L$. Moreover, in each ground state sector, 
four states were usually targeted. In most simulation, a single-site
 basis of 8 states (considering Rb particles as hard core)
is sufficient to faithfully represent a realistic system; this feature
has been systematically verified for small system sizes ($L\leq 12$),
and by using a 15 state single-site basis close to critical points.

As a case study, we considered fixed values of the optical lattice 
depth, and investigate the quantum phase diagram as a function
of the inter-species interaction strength. This choice is motivated by
the experimentally demonstrated tunability of Rb-K interactions\cite{catani2008},
which allows to span large interaction regimes by keeping the same
lattice structure. In Table \ref{tab:par}, we list the optical lattice setups 
investigated numerically, together with the corresponding 
hopping ratio. 

As already mentioned, PBCs assure very good scaling properties
during the finite size procedure, where we have always considered
polynomial scaling of the form:
\begin{equation}
\Delta_\alpha(1/L=x)=a_0+a_1x+a_2x^2.
\end{equation}
Different scaling {\it ans\"atze} with higher order contributions do not lead
to significant quantitative changes. Typical results are shown in Fig.~\ref{fig:gaps},
illustrating how both gaps opens for relatively small coupling strengths.

\begin{table}[t]  
    \centering
\begin{tabular}{||p{2cm}||*{5}{c|}|}
\hline
\bfseries       \centering{$s$}             &    20     &  20   &  26   &  26   \\
\hline
\bfseries      \centering{$s_\perp$}        &    50     &  80   &  70   &  90  \\
\hline  
    \centering{$t_K$ (nK)}                  &    5.4    &  5.4  &  2.8     &  2.8   \\
\hline  
    \centering{$t_{Rb}/t_K$}                &    0.050  &   0.050    &  0.033     & 0.033   \\
\hline
\end{tabular}
\caption{Optical lattice configurations investigated in this section, 
together with correspondent hopping asymmetry and typical
energy scales.\label{tab:par}}
\end{table}     
\begin{figure}\centering
\includegraphics[scale=0.28]{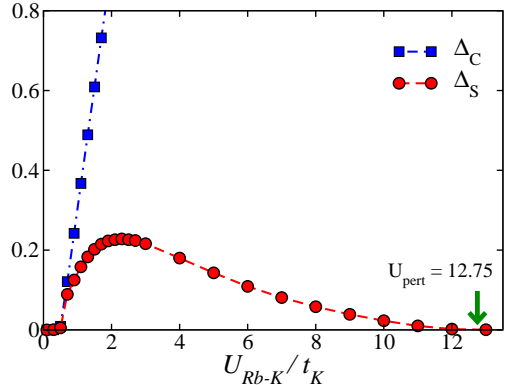}
\caption{Charge and spin gaps as a function of the inter particle interaction.
While the charge gap increases significantly and then saturates as a function 
of $U_{RbK}$, the spin gap displays a non-monotonic behavior; it open almost 
simultaneously as $\Delta_c$ (making the SCF region at small $U$ very thin),
displays a maximum around $U_{RbK}\simeq 2$, and then decreases till the 
BKT point separating the N\'eel phase from another SCF region. Numerical results
show a good agreement with second order perturbation theory estimate
(based on Eq.~\eqref{Jperp}, \eqref{Jz}), indicated
by a green arrow. } 
\label{fig:pd}
\end{figure}

For all values listed in Table~\ref{tab:par}, we observe the expected phase diagram,
which, as a function of $U_{RbK}$, presents a fully gapless phase first,
then an antiferromagnetic insulator, which melts into a SCF phase for 
relatively large values of the interaction strength. It is worth noticing 
that our procedure is not accurate enough to resolve a possible 
weak coupling SCF phase between fully gapless and AF phases;
since the gap is expected to open according to Berezinskii-Kosterlitz-Thouless scaling~\cite{gogolin_book}, refined methods should in principle be employed in order to 
quantitatively tackle this issue. Nonetheless, such SCF phase will
occupy a very small parameter region with extremely small charge gap,
making its experimental observation extremely challenging. 

Close to the SCF-AF critical point, we find excellent agreement between
numerical and perturbation theory results, as illustrated in Fig.~\ref{fig:pd};
we can then faithfully estimate the extent of the SCF region at strong 
coupling by using the results presented in Sec.~\ref{sec:3}. Depending on
the values of $s,s_\perp$, the SCF is present in a region which is relatively
wide, of order $U_{RbK}/t_K\simeq 2$, making its strong coupling observation
accessible in tunable systems. 

\begin{figure}
\centering
\includegraphics[scale=0.28]{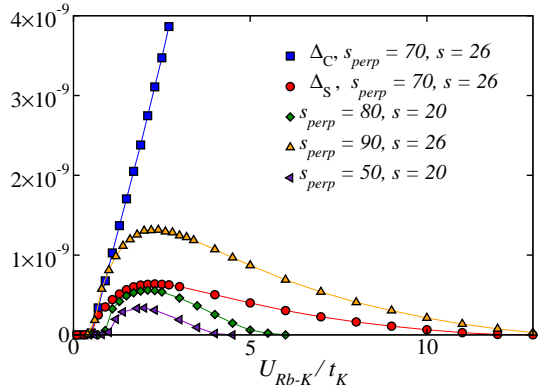}
\caption{Charge and spin gaps (expressed in K units) as a function of the inter 
particle interaction. The maximum value of the spin gap is always reached
far from the perturbative regime, making perturbation theory not fully 
reliable when estimating critical energy scales.
 } 
\label{fig:allgaps}
\end{figure}

A crucial point in the search for magnetic phases in ultra cold atomic
systems is the competition between thermal fluctuations and N\'eel order.
At a qualitative level, one expects that larger spin gaps could provide 
stronger finite-temperature signatures of ground state physics, since 
thermal fluctuations would then be not sufficiently strong to {\it activate}
spin excitations. In Fig.~\ref{fig:allgaps}, we plot the spin gap in 
temperature units (K) for several choices of the lattice configuration
as a function of the inter particle interactions.

Three main features emerge
from the quantitative analysis. The first one is, very deep lattices favor
larger values of $\Delta_s$, as expected, since they increase the 
ration between intra-species interactions and tunneling rates.
Secondly, we observe a quasi-plateau structure of the spin gap, that
is, its value is almost constant in a finite region of the AF phase;
this implies that inter-species interactions do not need accurate 
finite tuning in order to find the optimal experimental configuration.
Finally, it is worth noticing how the maximal value of $\Delta_s$
is usually far from the perturbative limit, signaling how strong 
coupling perturbation theory is usually not a fully trustable tool
in order to identify optimal experimental regimes where to 
search for AF. 

Finally, let us comment on the $\nu=2$ case. As suggested by 
the results discussed in the previous section, the topological Haldane phase
may emerge, but in very tiny parameter regimes; as such, it'd be interesting to 
investigate if such regimes, at least in the perturbative limit and eventually
via more accurate numerical techniques, may
enlarge in presence of larger mass imbalance by considering, e.g.,
$^6$Li-$^{133}$Cs mixtures. 

\section{Discussion and conclusions}\label{sec:concl}

We discussed magnetic phases in commensurate bosonic 
ultracold atomic gases, focusing on the recently realized
$^{87}$Rb-$^{41}$K mixtures in 1D optical lattices.
After an analysis on the relation between microscopic
and Hubbard parameters, we presented a combined analytical and numerical study
on the emergent insulating phases in such systems.
By means of perturbation theory, we 
illustrated the general magnetic scenarios accessible in such
setups by comparing effective spin-chain Hamiltonians with the
accessible parameter regimes, focusing on the relevant cases
of $S=1/2$ and $S=1$ spin representations.
In the latter case, we showed how, while the so called
{\it large-D} and ferromagnetic phases (phase separation)
are well within reach, exploring topological states of matter such
as the Haldane phase will represent a notable experimental
challenge. In the former, we discussed how the entire
strong coupling phase diagram, including ferromagnetic, 
antiferromagnetic, and super-counter-flow phases, may
be experimentally accessed by tuning the interspecies
repulsion by means of, e.g., a Feshbach resonance~\cite{catani2008}.
As a quantitative benchmark for experimental realizations, 
we have investigated the phase diagram of these systems
by means of DMRG simulations. By evaluating the typical
energy gaps in the systems, we estimated the optimal optical
lattice setups in order to observe anti-ferromagnetism 
in cold atomic gases, finding that experimentally challenging
but accessible temperatures~\cite{catani2009} in the nK range are required to
unambiguously reach such regimes.

\subsection*{Acknowledgements}

We would like to thank the entire LENS group, and in particular J. Catani, G. Lamporesi, and F. Minardi, 
for useful discussion on the experimental issues, and C. Degli Esposti Boschi, M. Di Dio and 
T. Roscilde for discussions on theoretical aspects. M.D. acknowledges support by
the European Commission via the integrated project AQUTE. This work was supported in part by INFN COM4 grant NA41.

\end{document}